\documentclass[graybox]{svmult}

\usepackage{type1cm}        
\usepackage{makeidx}         
\usepackage{graphicx}        
\usepackage{multicol}        
\usepackage[bottom]{footmisc}

\usepackage{newtxtext}       %

\usepackage{enumerate}
\usepackage{cite}
\usepackage{enumerate}
\usepackage{cite}
\usepackage{textcomp}
\usepackage{amsmath, amssymb, amsfonts}
\usepackage{mathrsfs}
\usepackage{mathtools}
\usepackage{dsfont}
\usepackage{accents}
\usepackage{color}
\usepackage{caption}

\DeclareMathOperator{\diag}{diag}
\DeclareMathOperator{\voi}{VoI}
\DeclareMathOperator{\EXP}{\mathsf{E}}
\DeclareMathOperator{\Cov}{\mathsf{cov}}
\DeclareMathOperator{\ProbM}{\mathsf{P}}
\DeclareMathOperator{\Prob}{\mathsf{p}}

\makeindex

\begin{document}

\renewcommand\footnotemark{}
\title*{Foundations of Value of Information: A Semantic Metric for Networked Control\\Systems Tasks \thanks{\hspace{-4.2mm}Corresponding Author: Touraj Soleymani (touraj@imperial.ac.uk). To be published as a book chapter by \it{Springer}.}}
\titlerunning{Foundations of Value of Information: A Semantic Metric}
\author{Touraj Soleymani, John S. Baras, Sandra Hirche, and Karl H. Johansson}
\institute{Touraj Soleymani \at University of London, United Kingdom
\and John S. Baras \at University of Maryland, United States
\and Sandra Hirche \at Technical University of Munich, Germany
\and Karl H. Johansson \at Royal Institute of Technology, Sweden}

\maketitle

\vspace{-10mm}
\abstract{In this chapter, we present our recent invention, i.e., the notion of the value of information---a semantic metric that is fundamental for networked control systems tasks. We begin our analysis by formulating a causal tradeoff between the packet rate and the regulation cost, with an encoder and a decoder as two distributed decision makers, and show that the valuation of information is conceivable and quantifiable grounded on this tradeoff. More precisely, we characterize an equilibrium, and quantify the value of information there as the variation in a value function with respect to a piece of sensory measurement that can be communicated from the encoder to the decoder at each time. We prove that, in feedback control of a dynamical process over a noiseless channel, the value of information is a function of the discrepancy between the state estimates at the encoder and the decoder, and that a data packet containing a sensory measurement at each time should be exchanged only if the value of information at that time is nonnegative. Finally, we prove that the characterized equilibrium is in fact globally optimal.}

\section{Introduction}
Semantic communication, which integrates the communication purpose directly into the design process, represents a radical transformation in the design of communication systems~\cite{uysal2022semantic}. This transformation is mainly motivated by the expansive domain of networked control systems, i.e., spatially distributed systems wherein feedback control loops are closed over communication channels~\cite{baillieul2007X, park2017, lee2023CAP}. Commonly, in a networked control system, data packets containing sensory measurements are transmitted to the controller in a periodic way as this facilitates the analysis of such a system~\cite{alur2007}. It has, however, been conceived that not every one of these data packets has the same effect on the system performance, and that one should employ a scheduler, aka event trigger, that transmits a data packet only when a significant deviation in the system occurs~\cite{Astrom:2002eg}. This adaptive communication has a major consequence: a dramatic reduction in the number of packet transmissions guaranteeing some level of system performance, which has been found appealing, and led to extensive development of event-triggered systems in different contexts even beyond control including signal processing~\cite{tsividis2010}, consensus~\cite{dimos2012}, optimization~\cite{meinel2014}, and fault detection~\cite{li2017}.

Although transmission of a data packet containing a sensory measurement in a networked control system decreases the uncertainty of the controller and improves the quality of regulation, it has indeed a price from the communication perspective. It is, therefore, rational that such a data packet be transmitted only if it is valuable in the sense of a cost-benefit analysis, i.e., only if its benefit surpasses its cost. Yet, the fact is that our knowledge has been very limited so far about this valuation of information and its connection with the above-mentioned adaptive communication. In the current chapter, we present our recent invention~\cite{touraj-thesis, voi, voi2}, which shed light on this intrinsic property of networked control systems, by formulating a causal tradeoff between the packet rate and the regulation cost. We show that the valuation of information is conceivable and quantifiable grounded on this tradeoff. We then argue that the value of information systematically captures the semantics of data packets by taking into account their potential impacts, and that a strategy based on the value of information optimally shapes the information flow in networked control systems. As such, the value of information can be regarded as a semantic metric~\cite{uysal2022semantic} that determines the right piece of information, a concept that is not defined in classical data communication, while it is crucial to the development of future communication~networks.

\subsection{Literature Survey}
In a networked control system, two distinct communication channels can be identified: observation channel and command channel. The former is a channel that links the sensor to the controller, while the later is a channel that links the controller to the actuator. Depending on the scenario of interest, a scheduler can be exploited at the sensor side to reduce the number of packet transmissions in the observation channel, or at the controller side to reduce that in the command channel. Our focus here lies in determining optimal decision policies in the setup where the scheduler and the controller operate separately as two distributed decision makers. Within this setup, {\AA}str\"om and Bernhardsson~\cite{Astrom:2002eg} showed that for a scalar linear diffusion process, with impulse control and under a sampling rate constraint, event-triggered sampling outperforms periodic sampling in the sense that it attains a lower mean error variance. 

There exist a number of studies, with restrictive assumptions, that are connected to our work concerning the determination of optimal decision policies~\cite{imer2010, lipsa2011, molin2017, chakravorty2016, rabi2012, guo2021-IT}. The intrinsic difficulty in these studies is due to the existence of two different information sets for the two decision makers, which complicates the derivation of the optimal policies. Notably, Imer and Ba{\c{s}}ar \cite{imer2010} studied the optimal event-triggered estimation of a scalar Gauss--Markov process based on dynamic programming by assuming that the scheduling policy is symmetric threshold, showed that the optimal estimator is linear, and derived the optimal threshold value. In addition, Rabi~\emph{et~al.}~\cite{rabi2012} formulated the optimal event-triggered estimation of the scalar Ornstein--Uhlenbeck process as an optimal multiple stopping time problem by assuming that the estimator is linear, and showed that the optimal scheduling policy is symmetric threshold. Lipsa and Martins~\cite{lipsa2011} analyzed the optimal event-triggered estimation of a scalar Gauss--Markov process based on majorization theory, and proved that the optimal scheduling policy is symmetric threshold and the optimal estimation policy is linear. Molin and Hirche~\cite{molin2017} studied the convergence properties of an iterative algorithm for the optimal event-triggered estimation of a scalar Markov process with symmetric noise distribution, and found a result coinciding with that~in~\cite{lipsa2011}.

Besides, several works have investigated optimal event-triggered estimation when the scheduling policy is entirely fixed~\cite{sijs2012, wu2013, he2018, han2015}. The main challenge in these works is to find a procedure for dealing with a signaling effect, which can cause a nonlinearity in the structure of the optimal estimator. To that end, Sijs and Lazar \cite{sijs2012} used a sum of Gaussian approximation, and developed an estimator that has an asymptotically bounded estimation error covariance for a Gauss--Markov process subject to a fixed deterministic scheduling policy. Wu~\emph{et~al.}~\cite{wu2013} used a Gaussian approximation, and found a suboptimal estimator for a Gauss--Markov process subject to a fixed deterministic threshold scheduling policy. He~\emph{et~al.}~\cite{he2018} took one step further, and adopted the generalized closed skew normal distribution to characterize the optimal estimator for a Gauss--Markov process subject to a similar scheduling policy. Han~\emph{et~al.}~\cite{han2015} also took advantage of a fixed stochastic scheduling policy that preserves the Gaussianity of the conditional distribution, and obtained the optimal estimator for a Gauss--Markov process.

Furthermore, several works have investigated optimal event-triggered control when the scheduling policy is entirely fixed~\cite{ramesh2013, molin2013, demirel2018}. Note that this problem is more complicated than the estimation counterpart because of a dual effect, which can lead to a coupling between estimation and control. In this context, Molin and Hirche~\cite{molin2013} studied the optimal event-triggered control of a Gauss--Markov process, and showed that the optimal control policy is certainty equivalent when the scheduling policy is reparametrizable in terms of primitive random variables. Ramesh~\emph{et~al.}~\cite{ramesh2013} studied the dual effect in the optimal event-triggered control of a Gauss--Markov process, and proved that the dual effect in general exists. They also proved that the certainty equivalence principle holds if and only if the scheduling policy is independent of the control policy. Later, Demirel~\emph{et~al.}~\cite{demirel2018} addressed the optimal event-triggered control of a Gauss--Markov process by adopting a stochastic scheduling policy that preserves the Gaussianity of the conditional distribution, and showed that the optimal control policy remains certainty equivalent.

\subsection{Contributions and Organization}
In this chapter, drawing on the work laid in~\cite{touraj-thesis, voi, voi2}, we explore the foundations of the value of information, and establish a theoretical framework for its quantification. More specifically, we first prove the existence of an equilibrium in a causal tradeoff between the packet rate and the regulation cost for a partially-observable multi-dimensional Gauss-Markov process without any limiting assumptions on the information structure or the policy structure, and quantify the value of information at this equilibrium, where the design of the encoder and the decoder becomes separated. Then, we prove that the characterized Nash equilibrium is in fact globally~optimal. We show that the value of information is a symmetric function of the discrepancy between the state estimates at the encoder and the decoder, and that a data packet containing a sensory measurement should be transmitted from the encoder to the decoder at each time only if the value of information at that time is nonnegative. Finally, we discuss that the value of information can be computed with arbitrary accuracy, and that it can be approximated by a closed-form quadratic function with a performance guarantee.

Note that, despite a considerable body of research in the area of networked systems, the characterization of the set of globally optimal solutions in the rate-regulation tradeoff, as described above, for multi-dimensional Gauss--Markov processes had been an open problem. We have characterized for the first time a policy profile that belongs to this set without imposing any restrictions on the information structure or the policy structure. We have proved that such a policy profile consists of a symmetric threshold scheduling policy and a certainty-equivalent control policy. More specifically, we have shown that the rate-regulation tradeoff attains a globally optimal solution of the form $(\pi^{\star},\mu^{\star}) = ( \{\mathds{1}_{\voi(k) \geq 0}\}_{k=0}^{N}, \{ - L(k) \hat{x}(k) \big\}_{k=0}^{N} )$, where $\mathds{1}_{\voi(k) \geq 0}$ denotes the indicator function of ${\voi(k) \geq 0}$, $\voi(k)$ is the value of information, $L(k)$ is the linear-quadratic-regulator gain, and $\hat{x}(k)$ is the minimum mean-square-error state estimate at the controller. Clearly, our study is different from the studies in~\cite{imer2010, lipsa2011, molin2017, chakravorty2016, rabi2012, guo2021-IT}, where the results apply to the estimation of scalar processes. Here, the results apply to the control of multi-dimensional Gauss--Markov processes. Our study is also different from the studies in~\cite{sijs2012, wu2013, han2015, he2018, molin2013, ramesh2013, demirel2018}, where an estimation policy or a control policy is derived when the scheduling policy is fixed and subject to some conditions. Here, we search for a globally optimal scheduling policy and a globally optimal control policy jointly and without any restrictions.

Traditionally, the value of information was defined as the value that can be assigned to the reduction of uncertainty from the decision maker's perspective, given a piece of information~\cite{howard1966}. This concept has found application across various disciplines, including information economics~\cite{stigler1961}, risk management~\cite{gould1974}, and stochastic programming~\cite{avriel1970}. Later, Dempster~\cite{dempster1981} and Davis~\cite{davis1989, davis1991} explored this concept in the context of optimal control. However, in \cite{davis1989, davis1991}, the value of information was defined as the variation in a value function with respect to relaxation of the non-anticipativity constraint at the controller. It is obvious that our perspective is fundamentally different from these previous works. We have transcended and redefined the notion of the value of information as a fundamental semantic metric applicable to networked control systems, cyber-physical systems, and distributed computational systems tasks.

The chapter is organized into 6 sections. The current introductory section is accompanied by Section~\ref{sec:prob-formulation}, where we formulate the causal tradeoff problem. We define formally the value of information in Section~\ref{sec:voi-def}. We discuss the existence of a signaling and dual effects in Sections~\ref{sec:signaling} and \ref{sec:dual}. Then, we present our main results on the characterization and the computation of the value of information in Sections~\ref{sec:emergence}~and~\ref{sec:global}. We propose a simple approximation of the value of information in Section~\ref{sec:approximation}. We provide a numerical example related to an inverted pendulum on a cart in Section~\ref{sec:examples}. Finally, we make concluding remarks in Section~\ref{sec:conclusion}.

\section{A Tradeoff in the Noiseless Communication Regime}\label{sec:prob-formulation}
Consider a basic networked control system composed of a dynamical process, a sensor with an encoder, an actuator with a decoder, and a noiseless channel that connects the sensor to the actuator. At each time~$k$, a message containing a new measurement, represented by~$\check{x}(k)$ can be transmitted over the channel from the sensor to the actuator, where an actuation input, represented by $u(k)$, should be computed causally in real time and over a finite time horizon~$N$. We assume that the time is discretized into time slots, and the duration of each time slot~is~constant.

The dynamical process is equipped with a sensor and an actuator that have computational capabilities. This process obeys the state and output equations
\begin{align}
	x(k+1) &= A(k) x(k) + B(k) u(k) + w(k)\label{eq:sys}\\[2\jot]
	y(k) &= C(k) x(k) + v(k) \label{eq:sens}
\end{align}
for $k \in \mathbb{N}_{[0,N]}$ with initial condition $x(0)$, where $x(k) \in \mathbb{R}^n$ is the state of the process, $A(k) \in \mathbb{R}^{n \times n}$ is the state matrix, $B(k) \in \mathbb{R}^{n \times m}$ is the input matrix, $u(k) \in \mathbb{R}^m$ is the actuation input, $w(k) \in \mathbb{R}^n$ is a Gaussian white noise with zero mean and covariance $W(k) \succ 0$, $y(k) \in \mathbb{R}^p$ is the output of the sensor, $C(k) \in \mathbb{R}^{p \times n}$ is the output matrix, and $v(k) \in \mathbb{R}^p$ is a Gaussian white noise with zero mean and covariance $V(k) \succ 0$. 

\begin{assumption}
For the dynamical process model in (\ref{eq:sys}) and (\ref{eq:sens}), the following assumptions are satisfied:
\begin{enumerate}[(i)]
	\item The initial condition $x(0)$ is a Gaussian vector with mean $m(0) \in \mathbb{R}^n$ and covariance $M(0) \succ 0$.
	\item The random variables $x(0)$, $w(t)$, and $v(s)$ for $t,s \in \mathbb{N}_{[0,T]}$ are mutually independent, i.e., $\Prob(x(0), w(0\!:\!T), v(0\!:\!T)) = \Prob(x(0)) \prod_{k=0}^{T} \Prob(w(k))$ $\times \prod_{k=0}^{T} \Prob(v(k))$.
\end{enumerate}
\end{assumption}

The communication channel between the sensor and the actuator is noiseless but costly, and the information in this channel is carried in the form of data packets. The input-output relation of this channel is given by
\begin{align}\label{eq:etm1}
z(k+1) = \left\{
  \begin{array}{l l}
     \check{x}(k), & \ \text{if} \ \sigma(k) =1, \\[1\jot]
     \mathfrak{E}, & \ \text{otherwise}
  \end{array} \right.
\end{align}
for $k \in \mathbb{N}_{[0,N]}$ with $z(0) = \mathfrak{E}$, where $z(k)$ is the output of the channel, $\mathfrak{E}$ represents absence of transmission, and $\sigma(k) \in \{0,1\}$ is the transmission decision decided by the encoder. The message that can be transmitted at time $k$ contains the minimum mean-square-error state estimate at the encoder at time $k$. Clearly, this state estimate condenses all the previous and current outputs of the dynamical process, and its transmission is always better than that of the raw output at time $k$.

\begin{assumption}
For the communication channel model in (\ref{eq:etm1}), the following assumptions are satisfied:
\begin{enumerate}[(i)]
	\item Quantization error is negligible, i.e., in a successful transmission, real-value sensory information can be conveyed from the encoder to the decoder without any error.
	\item A message sent at time $k$ is delivered without any loss or error to the decoder at time $k+1$.
\end{enumerate}
\end{assumption}

The encoder and the decoder, as two distributed decision makers, make their decisions at each time $k$ based on their causal information sets, which are given~by $\mathcal{I}(k) := \{ y(t), z(x), \sigma(s), u(s) | t \in \mathbb{N}_{[0,k]}, s \in \mathbb{N}_{[0,k-1]} \}$ and $\mathcal{J}(k) := \{ z(t), \sigma(s), u(s) | t \in \mathbb{N}_{[0,k]}, s \in \mathbb{N}_{[0,k-1]} \}$, respectively, We say that an encoding policy $\epsilon$ and a decoding policy $\delta$ are admissible if $\epsilon = \{\ProbM(\sigma(k) | \mathcal{I}(k)) \}_{k=0}^{N}$ and $\delta = \{\ProbM(u(k) | \mathcal{J}(k)) \}_{k=0}^{N}$, where $\ProbM(\sigma(k) | \mathcal{I}(k))$ and $\ProbM(u(k) | \mathcal{J}(k))$ are stochastic kernels defined on suitable measurable spaces. We represent the sets of admissible encoding policies and admissible decoding policies by $\mathcal{E}$ and $\mathcal{D}$, respectively. 

Note that the performance of a networked control system depends on the efficiency of communication and control subsystems. Therefore, we can enhance the overall system performance by jointly designing communication and control subsystems. Here, we are interested in a causal tradeoff between the packet rate and the regulation cost. The former penalizes the transmission of message in the channel, and is expressed as
\begin{equation}\label{eq:rate-measure}
R := \frac{1}{N+1} \EXP \bigg[\sum_{k=0}^{N} \ell(k) \sigma(k) \bigg],
\end{equation}
where $\ell(k) \geq 0$ is a weighting coefficient. However, the latter penalizes the state deviation and the control effort, and is expressed as
\begin{equation}\label{eq:control-measure}
\begin{aligned}
J:= \frac{1}{N+1} \EXP \bigg [\sum_{k=0}^{N+1} x(k)^T Q(k) x(k) + \sum_{k=0}^{N} u(k)^T R(k) u(k) \bigg]
\end{aligned}
\end{equation}
where $Q(k) \succeq 0$ and $R(k) \succ 0$ are weighting matrices. The causal tradeoff between the packet rate and the regulation cost can then be formulated as a two-player stochastic optimization problem with the loss function $\Phi := \lambda R + J$ over the space of admissible policy profiles $\mathcal{E} \times \mathcal{D}$ given a tradeoff multiplier $\lambda > 0$. This tradeoff, as we will see, allows us to describe the value of information.

\begin{remark}
The loss function $\Phi$ represents a tradeoff between two objective functions. The objective function in (\ref{eq:rate-measure}) is appropriate for packet switching networks. This objective function takes into account the price of communication through the weighting coefficient. Moreover, the objective function in (\ref{eq:control-measure}) is appropriate for regulation tasks. This objective function can be modified for tracking tasks by a transformation when the reference trajectory is known. Finally, note that the underlying optimization problem with the loss function $\Phi$ over the space of admissible policy profiles $\mathcal{E} \times \mathcal{D}$ is a decentralized stochastic problem, for which centralized stochastic methods cannot be applied directly. In this chapter, we seek an optimal solution, denoted by $(\epsilon^{\star},\delta^{\star})$, to this problem.
\end{remark}

\section{Quantification of the Value of Information}\label{sec:voi-def}
We should emphasize that the networked control system under study involves two decision makers with distinct information sets. Consequently, we can define two distinct value functions, viz., one from the perspective of the encoder, i.e., $V(k,\mathcal{I}(k))$, and one from the perspective of the decoder, i.e., $W(k,\mathcal{J}(k))$. Building upon this observation, we introduce our general formula of the value of information in the following definition.

\begin{definition}[Value of Information]
The value of information at time $k$ is defined as the variation in the value function $V(k,\mathcal{I}(k))$ with respect to the sensory measurement $\check{x}(k)$ that can be communicated from the encoder to the decoder at time~$k$,~i.e.,
\begin{align}\label{eq:voi-def}
\voi(k,\mathcal{I}(k)) := V(k,\mathcal{I}(k))|_{\sigma(k) = 0} - V(k,\mathcal{I}(k))|_{\sigma(k) = 1}
\end{align}
where $V(k,\mathcal{I}(k))|_{\sigma(k)}$ denotes the value function $V(k,\mathcal{I}(k))$ when the transmission decision $\sigma(k)$ is enforced.
\label{def:voi}
\end{definition}

\begin{remark}
The value of information $\voi(k,\mathcal{I}(k))$, defined in (\ref{eq:voi-def}), measures in a sense the sensitivity of the value function $V(k,\mathcal{I}(k))$ with respect to a data packet that can be exchanged at time $k$. Note that the above formula is general and valid for any choice of the system model. Furthermore, recall that we are interested in a valuation of information associated with a decision about transmission of a data packet through the observation channel at each time $k$. This decision is made by the encoder and according to the stochastic kernel $\ProbM(\sigma(k) | \mathcal{I}(k))$. For this reason, $\voi(k,\mathcal{I}(k))$ was evaluated based on the value function $V(k,\mathcal{I}(k))$, and not the value function $W(k,\mathcal{J}(k))$. Finally, note that the value of information can equivalently be expressed in terms of a state-action value function, i.e., $Q$ function, in the context of reinforcement learning.
\end{remark}

The next lemma introduces a loss function that is equivalent to the original loss function in the sense that it yields the same optimal decision policies. Associated with this loss function, we will subsequently delineate the value functions $V(k,\mathcal{I}(k))$ and $W(k,\mathcal{J}(k))$.

\begin{lemma}[\hspace{-0.01mm}\cite{stoccontrol}]
Let $S(k) \succeq 0$ be a matrix obeying the algebraic Riccati equation
\begin{equation}\label{eq:riccati}
\begin{aligned}
S(k) &= Q(k) + A(k)^T S(k+1) A(k) - A(k)^T S(k+1) B(k)\\[2.65\jot]
	&\qquad \quad \times (B(k)^T S(k+1) B(k) + R(k))^{-1} B(k)^T S(k+1) A(k)
\end{aligned}
\end{equation}
for $k \in \mathbb{N}_{[0,N]}$ with initial condition $S(N+1) = Q(N+1)$. Then,
\begin{align}\label{eq:phiprime}
	\Psi := \EXP \bigg[ \sum_{k=0}^{N} \theta(k) \sigma(k) + \eta(k) \bigg]
\end{align}
is equivalent to $\Phi$, where $\theta(k) = \ell(k) \lambda$, $\eta(k) =  (u(k) + L(k) x(k) )^T (B(k)^T S(k+1) B(k) + R(k) ) (u(k) + L(k) x(k))$, and $L(k) = (B(k)^T S(k+1) B(k) + R(k))^{-1} B(k)^T$ $S(k+1) A(k)$.
\label{lem:1}
\end{lemma}

\begin{definition}[Value functions]
The value functions $V(k,\mathcal{I}(k))$ and $W(k,\mathcal{J}(k))$ are defined~as
\begin{align}
	V(k,\mathcal{I}(k)) :=& \min_{\epsilon \in \mathcal{E} : \delta = \delta^\star}\EXP \Big[ \sum_{t=k}^{N} \theta(t) \sigma(t) + \eta(t+1) \Big| \mathcal{I}(k) \Big],\label{eq:Ve-def}\\[1.5\jot]
	W(k,\mathcal{J}(k)) :=& \min_{\delta \in \mathcal{D}: \epsilon = \epsilon^\star}\EXP \Big[ \sum_{t=k}^{N} \theta(t-1) \sigma(t-1) + \eta(t) \Big| \mathcal{J}(k) \Big],\label{eq:Vc-def}
\end{align}
for $k \in \mathbb{N}_{[0,N]}$ given a policy profile $(\epsilon^{\star},\delta^{\star})$, where we adopt the convention $\theta(-1) = 0$, $\eta(N+1) = 0$, and $S(N+2) = 0$.
\end{definition}

\section{Existence of a Signaling Effect}\label{sec:signaling}
Note that events related to the absence of transmission, denoted as $z(k) = \mathfrak{E}$, provide an opportunity for the encoder to implicitly encode additional information beyond the explicitly communicated sensory measurements through successful packet deliveries. In this section, we will observe that this phenomenon, which is referred to as signaling effect, can directly impact the uncertainty of the decoder. In particular, we will see how the structure of the optimal estimator at the decoder is influenced by the structure of the scheduling policy used by the~encoder. 

Let $\check{x}(k)$ and $\hat{x}(k)$ denote the minimum mean-square-error state estimates at the encoder and the decoder at time $k$, respectively. Define the estimation error from the perspective of the encoder $\check{e}(k) := x(k) - \check{x}(k)$, the estimation error from the perspective of the decoder $\hat{e}(k) := x(k) - \hat{x}(k)$, and the estimation mismatch $\tilde{e}(k) := \check{x}(k) - \hat{x}(k)$. The following two Lemmas characterize the optimal estimators at the encoder and the decoder.

\begin{lemma}[\hspace{-0.01mm}\cite{stoccontrol}]
The conditional mean $\EXP[{x}(k) | \mathcal{I}(k)]$ is the minimum mean-square-error estimator at the encoder, and obeys
\begin{align}
	\check{x}(k+1) &= A(k) \check{x}(k) + B(k) u(k) \nonumber\\[2.45\jot]
	&\quad + K(k+1) \big(y(k+1) - C(k+1) ( A(k) \check{x}(k) + B(k) u(k))\big) \label{eq:kf-E} \\[2.45\jot]
	O(k+1) &= \big( (A(k) Y(k) A(k)^T + W(k))^{-1} \nonumber\\[2\jot]
	&\qquad \qquad \qquad \qquad \qquad + C(k+1)^T V(k+1)^{-1} C(k+1) \big)^{-1} \label{eq:kf-Cov}
\end{align}
for $k \in \mathbb{N}_{[0,N]}$ with initial conditions $\check{x}(0) = m(0) + O(0) C(0)^T V(0)^{-1}(y(0) - C(0) m(0))$ and $O(0) = (M(0)^{-1} + C(0)^T V(0)^{-1} C(0))^{-1}$, where $\check{x}(k) = \EXP[{x}(k) | \mathcal{I}(k)]$, $O(k) = \Cov[x(k) | \mathcal{I}(k)]$, and $K(k) = O(k) C(k)^T V(k)^{-1}$.
\label{c1:prop:imperf-KF}
\end{lemma}

\begin{lemma}[\hspace{-0.01mm}\cite{voi}]
The conditional mean $\EXP[{x}(k) | \mathcal{J}(k)]$ is the minimum mean-square-error estimator at the decoder, and obeys
\begin{align}\label{c1:eq:imperf-estimate-dynX}
	\hat{x}(k+1) &= A(k) \hat{x}(k) + B(k) u(k) + \sigma(k) A(k) \tilde{e}(k) + (1-\sigma(k)) \imath(k)
\end{align}
for $k \in \mathbb{N}_{[0,N]}$ with initial condition $\hat{x}(0) = m(0)$, where $\hat{x}(k) = \EXP[{x}(k) | \mathcal{J}(k)]$ and $\imath(k) = A(k) \EXP[\hat{e}(k) | \mathcal{J}(k),\sigma(k)=0]$. In addition, the conditional covariance $\Cov[{x}(k) | \mathcal{J}(k)]$ obeys
\begin{align}\label{c1:eq:imperf-cov-dynX}
	E(k+1) &= A(k) E(k) A(k)^T + W(k) \nonumber\\[2.05\jot]
	&\qquad \quad \ - \sigma(k) A(k) (E(k) - O(k)) A(k)^T - (1-\sigma(k)) \Xi(k),
\end{align}
for $k \in \mathbb{N}_{[0,N]}$ with initial condition $Z(0) = M(0)$, where $E(k) = \Cov[x(k) | \mathcal{J}(k)]$ and $\Xi(k) = A(k) (E(k) - \Cov[\hat{e}(k) | \mathcal{J}(k), \sigma(k) =0]) A(k)^T$.
\label{c1:prop:imperf-estimatorX}
\end{lemma}

\begin{remark}
Note that the optimal estimators at the encoder and the decoder have completely different characteristics. While the conditional distribution $\ProbM(x(k) | \mathcal{I}(k))$ is Gaussian and the conditional mean $\check{x}(k)$ obeys a linear recursive equation, the conditional distribution $\ProbM(x(k) | \mathcal{J}(k))$ is generally non-Gaussian and the conditional mean $\hat{x}(k)$ generally obeys a nonlinear recursive equation. Moreover, note that the residuals $\imath(k)$ and $\Xi(k)$ in (\ref{c1:eq:imperf-estimate-dynX}) and (\ref{c1:eq:imperf-cov-dynX}) are both due to implicit information. The existence of these terms implies that the decoder might be able to decrease its uncertainty even when it does not receive any data packet from the sensor. The values of the residuals $\imath(k)$ and $\Xi(k)$ at each time $k$ depend on the structure of the scheduling policy. For any fixed scheduling policy, these values can be computed numerically by techniques from nonlinear filtering (see e.g., \cite{sijs2012,wu2013}).
\end{remark}

\section{Existence of a Dual Effect}\label{sec:dual}
Note that the decoder (i.e., controller) in addition to affecting the state of the process under control might be able to decrease the future uncertainty of the state of the process. The existence of this phenomenon, which is referred to as dual effect, complicates the underlying problem. The reason is that such a dual role of the controller leads to a coupling between estimation and control, a complicated situation where the separation principle no longer holds. The dual effect of the controller has formally defined in the next definition.

\begin{definition}[Dual effect]
For a given feedback control system, let $\mathcal{J}(k)$ be the information set of the controller at time $k$, and $\tilde{\mathcal{J}}(k)$ be the information set of the controller at time $k$ when all actuation inputs are equal to zero. The control has no dual effect of order $r$, $r \geq 2$, if
\begin{align*}
	\EXP[M_{i}^r(k) | \mathcal{J}(k) ] = \EXP[M_{i}^r(k) | \tilde{\mathcal{J}}(k)],
\end{align*}
where $M_{i}^r(k) = (x_{i}(k) - \EXP[x_{i}(k) | \mathcal{J}(k)])^r$ is the $r$th central moment of the $i$th component of the state $x(k)$ conditioned on $\mathcal{J}(k)$. In other words, the control has no dual effect if the expected future uncertainty is not affected by the prior actuation~inputs.
\end{definition}

The next lemma shows the existence of a dual effect in the networked control system under study.

\begin{lemma}[\hspace{-0.01mm}\cite{ramesh2013}]
The causal tradeoff between the packet rate and the regulation cost is in general subject to a dual effect of order $r = 2$.
\end{lemma}

\begin{remark}
The existence of the dual effect in the contexts of stochastic control and event-triggered control has been demonstrated in \cite{bar1974dual, ramesh2013}. Notably, in \cite{ramesh2013}, the authors proved this phenomenon by considering a general scheduling policy that depends on the state of the dynamical process, and searching for the optimal control policy. They showed that, for a linear system, the optimal control policy in fact becomes a nonlinear policy, with no close-form solutions. Note that, in general, addressing the challenges posed by the dual effect remains an ongoing endeavor, necessitating development of advanced control and estimation techniques.
\end{remark}

\section{Emergence of the Value of Information}\label{sec:emergence}
In this section, we show how the value of information emerges from the causal tradeoff formulated in Section~\ref{sec:prob-formulation}. More specifically, we will shed light on the structure of the value of information, and show that the optimal policies hinge on this quantity. Note that a solution concept that can be defined associated with the causal tradeoff between the packet rate and the regulation cost with two decision makers is Nash equilibria, which is captured by in the following definition.

\begin{definition}[Nash equilibrium]
For a given team game with two decision makers, let $\gamma^1 \in \mathcal{G}^1$ and $\gamma^2 \in \mathcal{G}^2$ be the decision policies of the decision makers, where $\mathcal{G}^1$ and $\mathcal{G}^2$ are the sets of admissible policies, and $\Omega(\gamma^1,\gamma^2)$ be the associated loss function. A policy profile $(\gamma^{1\star},\gamma^{2\star})$ represents a Nash equilibrium if
\begin{align*}
	\Omega(\gamma^{1\star},\gamma^{2\star}) \leq \Omega(\gamma^{1},\gamma^{2\star}), \ \text{for all } \gamma^1 \in \mathcal{G}^1,\\[2.25\jot]
	\Omega(\gamma^{1\star},\gamma^{2\star}) \leq \Omega(\gamma^{1\star},\gamma^{2}), \ \text{for all } \gamma^2 \in \mathcal{G}^2.
\end{align*}
\end{definition}
Note that at a Nash equilibrium the value functions $V(k,\mathcal{I}(k))$ and $W(k,\mathcal{J}(k))$ should simultaneously satisfy the optimality relations (see e.g., \cite{basargame}).

Define $\varrho(k) = \EXP[V(k+1,\mathcal{I}(k+1))|\mathcal{I}(k), \sigma(k) = 0] - \EXP[V(k+1,\mathcal{I}(k+1))|\mathcal{I}(k), \sigma(k) = 1]$ and $\Gamma(k) = A(k)^T S(k+1) B(k) (B(k)^T S(k+1) B(k) + R(k))^{-1} B(k)^T S(k+1) A(k)$. In the next theorem, we characterize a Nash equilibrium in the causal tradeoff between the packet and regulation cost.
\begin{theorem}[\hspace{-0.01mm}\cite{voi}]
The causal tradeoff between the packet rate and the regulation cost pertaining to feedback control of a partially observable process modeled by (\ref{eq:sys}) and (\ref{eq:sens}) over a noiseless channel modeled by (\ref{eq:etm1}) admits at least one Nash equilibrium $(\epsilon^{\star},\delta^{\star})$ such that
\begin{align}\label{eq:opt-profile}
(\epsilon^{\star},\delta^{\star}) = \Big( \big\{\mathds{1}_{\voi(k,\mathcal{I}(k)) \geq 0} \big\}_{k=0}^{N}, \big\{ - L(k) \hat{x}(k) \big\}_{k=0}^{N} \Big),
\end{align}
with the value of information $\voi(k,\mathcal{I}(k))$ as a symmetric function of the estimation mismatch $\tilde{e}(k)$ obeying
\begin{align}
	\voi(k, \mathcal{I}(k)) = \tilde{e}(k)^T A(k)^T \Gamma(k+1) A(k) \tilde{e}(k) - \theta(k) + \varrho(k),\label{eq:voi-imperfectX}
\end{align}
and with the conditional mean $\hat{x}(k)$ as a state estimation unaffected by implicit information obeying
\begin{align}
\hat{x}(k+1) = A(k) \hat{x}(k) + B(k) u(k) + \sigma(k) A(k) \tilde{e}(k),
\end{align}
for $k \in \mathbb{N}_{[0,N]}$ with initial condition $\hat{x}(0) = m(0)$.
\label{c1:thm:imperf-optimalityX}
\end{theorem}

\begin{remark}
Our structural result shows that at the equilibrium $(\epsilon^\star,\delta^\star)$ the design of the encoder and the decoder in (\ref{eq:opt-profile}) becomes separated, the optimal estimator at the decoder in (\ref{c1:eq:imperf-estimate-dynX}) becomes linear, and the conditional covariance in (\ref{c1:eq:imperf-cov-dynX}) becomes independent of the previous control inputs, implying that the control has no dual effect. In addition, our result shows that $\voi(k, \mathcal{I}(k))$ is a symmetric function of the estimation mismatch $\tilde{e}(k)$, and that it can be computed with arbitrary accuracy through solving the optimality equation recursively and backward in time. The complexity of this computation is $\mathcal{O}(N d^n s)$ when the estimation mismatch $\tilde{e}(k)$ is discretized in a grid with $d^{n}$ points and the associated expected value is obtained based on a weighted sum of $s$ samples.
\end{remark}

\begin{remark}
We argue that instead of fixing an ad-hoc triggering condition and studying the properties of the resulting event-triggered system, i.e., the procedure that has been used in most of the studies on event-triggered estimation and control, one should study a cost-benefit analysis without any limiting assumptions on the information structure or the policy structure, and find a triggering condition as a result of this analysis. Note that $\voi(k, \mathcal{I}(k))$ is in fact the difference between the benefit and the cost of a data packet. In light of our structural result, at each time $k$, the benefit of transmitting a data packet is $\tilde{e}(k)^T A(k)^T \Gamma(k+1) A(k) \tilde{e}(k) + \varrho(k)$ and its associated cost is $\theta(k)$. In this respect, our triggering condition has an important interpretation: a data packet containing the sensory measurement $\check{x}(k)$ should be transmitted to the decoder only if its benefit surpasses its cost, i.e., $\voi(k, \mathcal{I}(k)) \geq 0$. This interpretation does not exist for any triggering condition that is not based on a cost-benefit analysis.
\end{remark}

\section{Global Optimality Analysis}\label{sec:global}
Note that the tradeoff between the packet rate and the regulation cost in our study might admit multiple Nash equilibria. Unfortunately, there exists no general procedure for finding all these equilibria (if any). Using  backward induction, we proved the existence of a Nash equilibrium $(\epsilon^\star,\delta^\star)$, which has desirable characteristics. Our result guarantees that the set of globally optimal solutions cannot be empty. A natural question that arises in relation to the equilibrium $(\epsilon^\star,\delta^\star)$ is whether it is globally optimal. We study this issue in this section.

\begin{definition}[Globally optimal solutions]
For a given team game with two decision makers, let $\gamma^1 \in \mathcal{G}^1$ and $\gamma^2 \in \mathcal{G}^2$ be the decision policies of the decision makers, where $\mathcal{G}^1$ and $\mathcal{G}^2$ are the sets of admissible policies, and $\Omega(\gamma^1,\gamma^2)$ be the associated loss function. A policy profile $(\gamma^{1\star},\gamma^{2\star})$ is globally optimal if
\begin{align*}
	\Omega(\gamma^{1\star},\gamma^{2\star}) \leq \Omega(\gamma^{1},\gamma^{2}), \ \text{for all } \gamma^1 \in \mathcal{G}^1, \gamma^2 \in \mathcal{G}^2.
\end{align*}
\end{definition}

\begin{theorem}[\hspace{-0.01mm}\cite{voi2}]
The equilibrium $(\epsilon^{\star},\delta^{\star})$ characterized in Theorem 1 is globally optimal.
\label{thm:2}
\end{theorem}

\begin{remark}
This result is important as it shows that the characterized Nash equilibrium in~\cite{voi} has zero optimality gap. Observe that the decision policies $\epsilon^{\star}$ and $\delta^{\star}$ are deterministic, implying that randomization does not improve the system performance, and that they can be designed separately. Moreover, note that $\voi(k,\mathcal{I}(k))$ is a symmetric function of the estimation mismatch $\tilde{e}(k)$. According to these results, it is globally optimal that a data packet containing the sensory measurement $\check{x}(k)$ be transmitted to the decoder only if its benefit surpasses its cost, i.e., $\voi(k,\mathcal{I}(k)) \geq 0$. Finally, we remark that at the globally optimal solution $(\epsilon^\star,\delta^\star)$ the transmission of the state estimate $\check{x}(k)$ is equivalent to that of the estimation mismatch $\tilde{e}(k)$, whose magnitude is comparatively smaller.
\end{remark}

\section{Approximation of the Value of Information}\label{sec:approximation}
The computation of $\voi(k,\mathcal{I}(k))$, which requires solving the optimality equation associated with $V(k,\mathcal{I}(k))$, can be difficult especially when $n$ increases. This motivates us to search for an approximation of the value of information that can be expressed analytically. The next proposition provides such an approximation with a performance guarantee.

\begin{proposition}[\hspace{-0.01mm}\cite{voi}]
Let the control policy $\delta^{\star}$ be fixed. A scheduling policy $\epsilon^+$ that outperforms the periodic scheduling policy with period one in the rate-regulation tradeoff is given by
\begin{align}
\sigma(k)^+ = \mathds{1}_{\voi^{+}(k,\mathcal{I}(k)) \geq 0},
\end{align}
where $\voi^{+}(k,\mathcal{I}(k))$ is a quadratic approximation of the value of information expressed~as
\begin{align}\label{eq:voi-approx-2}
	\voi^{+}(k,\mathcal{I}(k)) = \tilde{e}(k)^T A(k)^T \Gamma(k+1) A(k) \tilde{e}(k) - \theta(k).
\end{align}
\label{c1:prop-Delta-imperfX}
\end{proposition}
\vspace{-0.6cm}

\begin{remark}
The value of information approximate $\voi^{+}(k, \mathcal{I}(k))$ is a closed-form quadratic function of the estimation mismatch $\tilde{e}(k)$, which does not depend on the cost-to-go terms. Our result provides a performance guarantee for this approximation in the sense that the scheduling policy $\epsilon^+$ synthesized based on $\voi^{+}(k, \mathcal{I}(k))$ outperforms the periodic scheduling policy with period one, when the certainty-equivalent control policy $\delta^\star$ is used. The result is obtained by exploiting a rollout algorithm, which can be viewed as a single iteration of the method of policy iteration.
\end{remark}

\begin{figure}[t]
\center
\vspace{-2.5mm}
  \includegraphics[width= 0.83\linewidth]{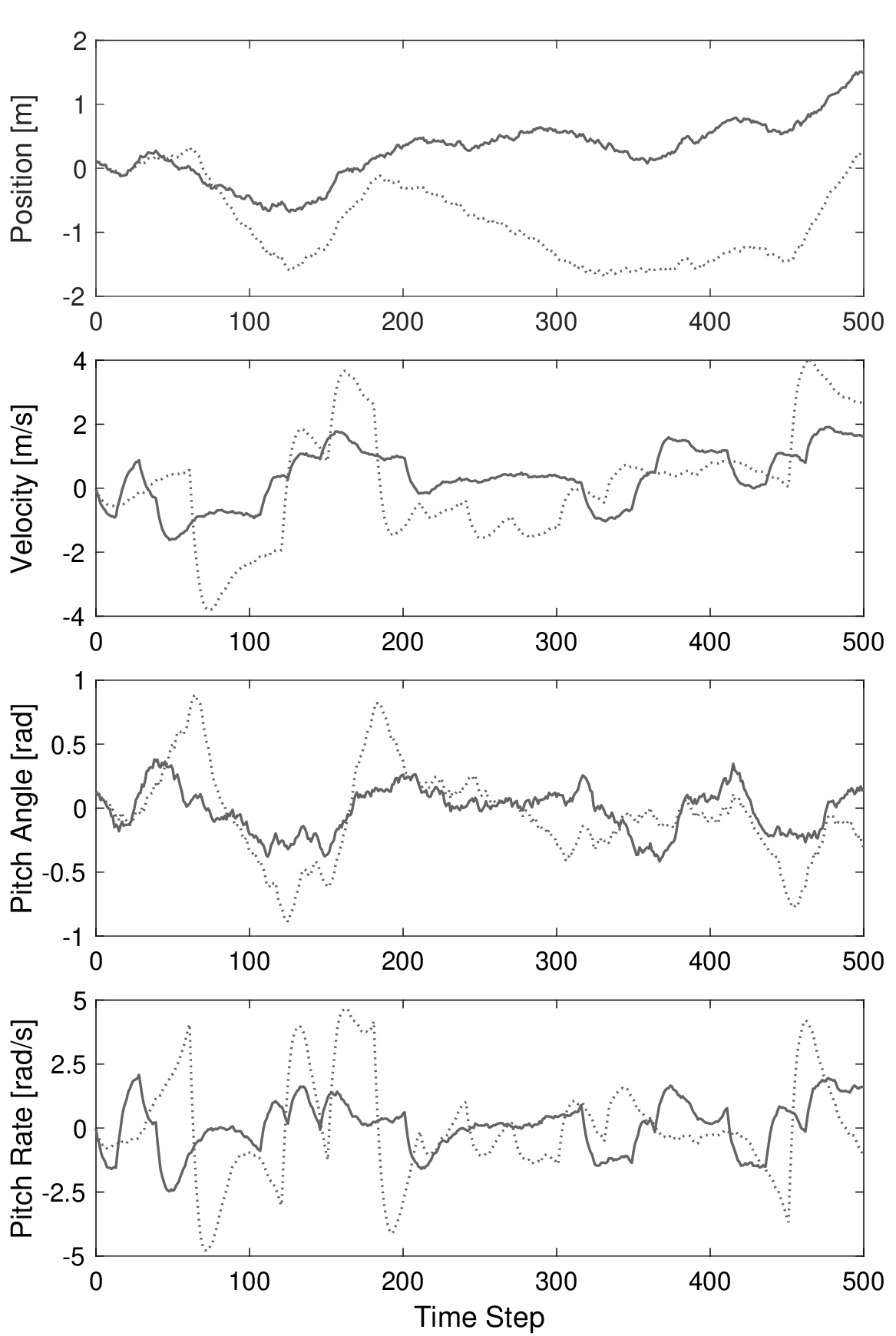}
  \caption{The value of information, transmission decision, and control input trajectories. The value of information is scaled by one tenth. The solid lines represent the trajectories under the scheduling policy designed based on the value of information, and the dotted lines represent the trajectories under a periodic scheduling policy.}
  \label{c1:fig:trajectories2X}
\end{figure}

\begin{figure}[t]	
\center
\vspace{-2.5mm}
  \includegraphics[width= 0.83\linewidth]{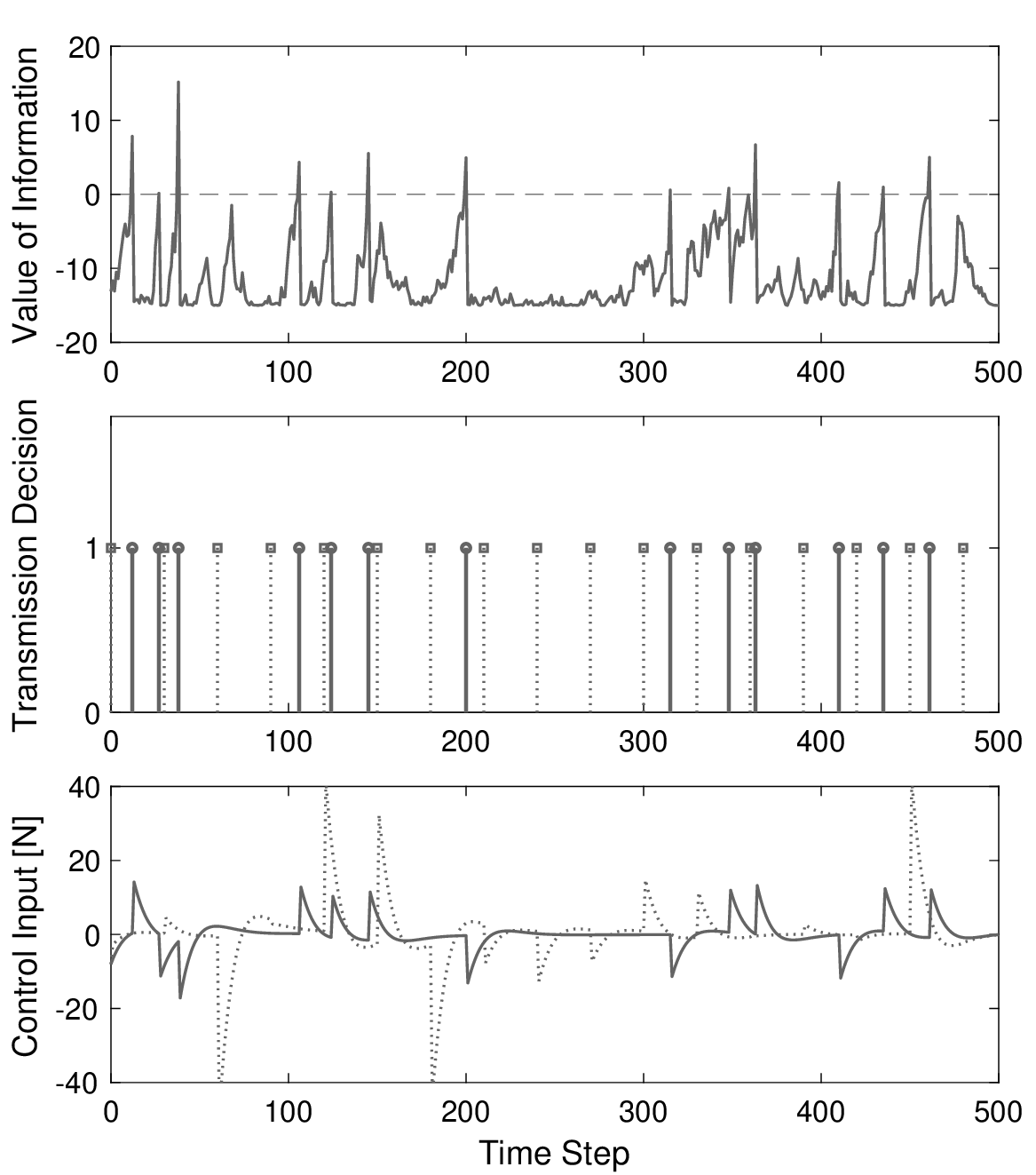}
  \caption{The position, velocity, pitch angle, and pitch rate trajectories. The solid lines represent the trajectories under the scheduling policy designed based on the value of information, and the dotted lines represent the trajectories under a periodic scheduling policy.}
  \label{c1:fig:trajectories1X}
\end{figure}

\section{Numerical Example}\label{sec:examples}
In this section, we provide a numerical example that demonstrates our theoretical results. We consider an inverted pendulum on a cart, for which the continuous-time equations of motion linearized around the unstable equilibrium are given~by
\begin{align*}
	(M+m) \ddot{x} + b \dot{x} - m l \ddot{\phi} = u\\[2.25\jot]
	(I + m l^2) \ddot{\phi} - m g l \phi = m l \ddot{x}
\end{align*}
where $x$ is the position of the cart, $\phi$ is the pitch angle of the pendulum, $u$ is the force applied to the cart, $M = 0.5 \ \text{kg}$ is the mass of the cart, $m = 0.2 \ \text{kg}$ is the mass of the pendulum, $b = 0.1 \ \text{N/m/sec}$ is the coefficient of friction for the cart, $l = 0.3 \ \text{m}$ is the distance from the pivot to the pendulum's center of mass, $I = 0.006 \ \text{kg.m$^2$}$ is the moment of inertia of the pendulum, and $g = 9.81 \ \text{m/s$^2$}$ is the gravity. The discrete-time state equation of the form~(\ref{eq:sys}) is specified by $A(k) = [1, 0.01, 0.0001, 0; 0, 0.9982, 0.0267, 0.0001; 0, 0, 1.0016, 0.01; 0, -0.0045, 0.3122,$ $1.0016]$, $B(k) = [0.0001; 0.0182; 0.0002; 0.0454]$, $W(k) = [0.0006, 0.0003, 0.0001,$ $0.0006; \ 0.0003, 0.0008, 0.0003, 0.0004;\ 0.0001, 0.0003, 0.0007, 0.0006; \ 0.0006,$\\$0.0004, 0.0006, 0.0031]$ for $k \in \mathbb{N}_{[0,N]}$, with the mean and the covariance of the initial condition $m(0) = [0; 0; 0.2;0]$ and $M(0) = 10W(k)$. Suppose that a sensor measures the position and the pitch angle at each time. The output equation of the form (\ref{eq:sens}) is specified by $C(k) =[1, 0, 0, 0; 0, 0, 1, 0]$ and $V(k) = [0.002, 0; 0, 0.001]$ for $k \in \mathbb{N}_{[0,N]}$. We are interested in a tradeoff between the packet rate and the regulation cost. The loss function $\Phi$ is specified by $Q(N+1)= \diag\{1,1,1000,1\}$, $\ell(k) = 1$, $Q(k)= \diag\{1,1,1000,1\}$, and $R(k)=1$ for $k \in \mathbb{N}_{[0,N]}$, time horizon $N= 500$, and tradeoff multiplier $\lambda = 0.0066$.

For a realization of this system, the value of information, transmission decision, and control input trajectories are shown in Fig.~\ref{c1:fig:trajectories2X}, and the position, velocity, pitch angle, and pitch rate trajectories in Fig.~\ref{c1:fig:trajectories1X}. Note that in this experiment, the value of information became nonnegative only $13$ times, which led to the transmission of a data packet from the sensor to the decoder at each of those times. The corresponding trajectories under a periodic scheduling policy with a higher number of transmissions are also illustrated in Fig.~\ref{c1:fig:trajectories2X} and \ref{c1:fig:trajectories1X}. We observe that the system under the scheduling policy designed based on the value of information was able to achieve relatively much better regulation quality.

\section{Conclusion}\label{sec:conclusion}
In this chapter, we introduced the notion of the value of information as an intrinsic property of networked control systems, and established a theoretical framework for its characterization and computation. The results asserted that the value of information systematically measures the semantics of each data packet as the difference between its benefit and its cost, and that a strategy based on the value of information optimally manages the communication between the sensor and the actuator by allowing only data packets with nonnegative valuations to be transmitted. Note that the above objectives could not be achieved by means of the traditional information-theoretic metrics or the traditional event-triggered conditions.

\bibliography{../../../../mybib}
\bibliographystyle{ieeetr}

\end{document}